%% file: cos_talk_martinez_1.tex
\documentclass[vecphys]{svmult}

\usepackage{makeidx}         
\usepackage{graphicx}        
\usepackage{multicol}        
\usepackage[bottom]{footmisc}

\makeindex             

\begin{document}

\title*{Nuclear Activity in UZC Compact Groups of Galaxies}

\author{M.A. Martinez\inst{1}, A. del Olmo\inst{1}, P. Focardi\inst{2}
  \and J. Perea\inst{1} }

\institute{Instituto de Astrof\'{\i}sica de Andaluc\'{\i}a (IAA) CSIC
\texttt{geli@iaa.es,chony@iaa.es,jaime@iaa.es}
\and Dipartamento di Astronomia-Universit\'a di Bologna (Italia)\\
\texttt{paola.focardi@unibo.it}}

\maketitle
%

\section{Introduction}
\label{sec:1}
It is well known, and accepted, that environment can play an important
role on galaxy formation and evolution, in particular it is expected
to influence mass assembly, star formation, morphological evolution
and even induce AGN phenomena. Several questions remain, however, open
as observational data still lack statistical robustness and give
somewhat conflicting results.  To properly assess and quantify the
environment effect on activation of nuclear activity (AGN and
Starburst) in galaxies we have chosen the galaxy systems which have
high galaxy density and low velocity dispersion, conditions that
maximize the number of gravitational encounters. Compact Groups of
Galaxies (CGs) satisfy both conditions and, moreover, being located in
low density surrounding environments, guarantee that environment
influence may arise only from the galaxy interactions within the CG.

To carry out this study we have chosen the sample of Compact Groups
(UZC-CGs) selected by \cite{Focardi02} applying an objective neighbour
search algorithm to UZC catalogue \cite{Falco99} ( Falco E.E., Kurtz,
M.J., Geller M.J. et al. 1999, PASP 111, 438).  The sample is
complete, large (986 galaxies in 291 CGs) and homogeneous and thus
particularly suited for our aim. We have, thus, collected available
spectroscopic data for UZC-CG galaxies.

\section{Archive Data}
\label{sec:2}

We have inspected three database archives looking for UZC-CG galaxies
spectra.  The databases are the Z-Machine Spectrograph Archive, the
FAST Spectrograph Archive and the Sloan Digital Sky Survey DR4.  We
have found 652, 246 and 221 spectra respectively. In several cases the
spectrum of the same galaxy was available in more than one Archive,
thus the total number of UZC-CG galaxies with spectrum is 868. The
vast majority of UZC-CGs (215 CGs) has spectroscopic full coverage,
i.e all member galaxies have a spectrum, and constitues our Complete
(C) sample. UZC-CGs with spectra for more than half member galaxies
are part of the Roughly Complete (RC) sample and UZC-CGs with less
than half member galaxies with spectra are parte of the Incomplete (I)
sample.  Table \ref{table1} summarizes the spectroscopical coverage of
the UZC-CG sample giving also information on the number of spectra
with emission feautures in each sample (column 5). However, for sake
of completness, in our study we have made use only of the Complete
sample.

\begin{table}
\centering
\caption{Available spectra}
\label{table1}    
\begin{tabular}{ccccc}
\hline\noalign{\smallskip}
Sample~& ~Groups~ & ~Galaxies~ & ~Spectra~ & ~Emissions~  \\
\noalign{\smallskip}\hline\noalign{\smallskip}

UZC-CG& 291 &  986 &  868  &  543  \\
C     & 215 &  720 &  720  &  485  \\
RC    & 47  &  171 &  120  &  75   \\
I     & 29  &  95  &  28   &  20   \\
 
\noalign{\smallskip}\hline
\end{tabular}
\end{table}

Since we do not have homogeneous spectroscopic data we have checked if
equivalent width (EW) provided by the three databases are similar.

\begin{figure}
\centering
\includegraphics[width=6cm,angle=-90]{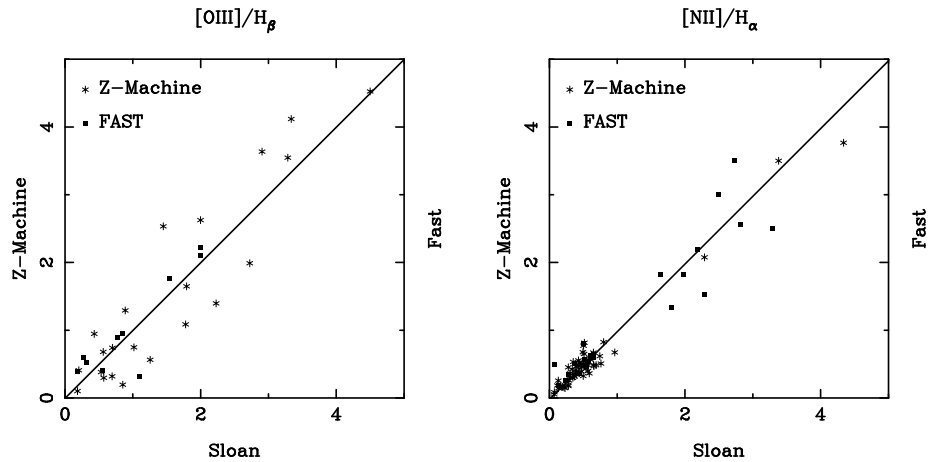}
\caption{The equivalent width line ratios in the three Archive data. Solid
line corresponds to slope 1 indicating the equivalence between the 
measurements}
\label{fig1}
\end{figure}

To do that, we have used the spectra, available for the same galaxies,
in the Z-Machine and Fast Spectrograph Archives and Sloan. From Fig
\ref{fig1} we can see that EW([OIII])/EW(H$_{\beta}$) and
EW([NII])/EW(H$_{\alpha}$) ratios are equivalent in the three Archive.

We have visually inspected all the spectra looking for the following
emission lines: H$_{\beta}$, [OIII] 4959\AA, 5007\AA, H$_{\alpha}$,
[NII]6548\AA, 6583\AA, the two sulphur [SII]6717\AA, 6731\AA\ . When
the EW were not available in the databases we have measured them,
using SPLOT task in IRAF. In cases of high S/N we have measured also
the EW of [OI] 6300\AA.

Emission line detection depends on spectral quality, in fact the
fraction of UZC-CG galaxies with emission line spectra is 58 \% for
spectra from the Z-Machine Archive, 53 \% for spectra from the Fast
Archive and 80\% for spectra from the Sloan.  Since the majority of
spectra for our sample comes from the Z-Machine Archive we can safely
state that the fraction of emission line spectra in our sample must be
considered a lower limit estimate.

\section{The Sample}
\begin{figure}
\centering
\includegraphics[width=4cm,angle=-90]{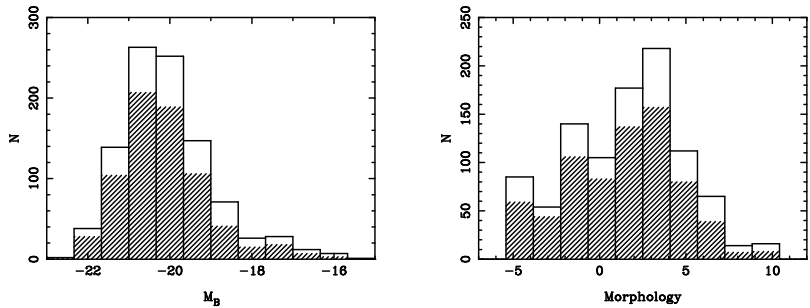}
\caption{Absolute magnitude and morphology distributions for the whole
UZC-CG catalogue (solid line) and for our sample of Complete groups
 (dashed histogram).}
\label{fig2}
\end{figure}

Our Complete sample consist of 215 UZC-CGs (720 galaxies). Figure
\ref{fig2} shows that it is well representative of the whole UZC-CG
sample. In fact there is not significant difference between the
Complete sample (dashed diagram) and the whole UZC-CG sample(solid
line), neither in absolute magnitude nor in morphology. (B magnitude
and morphology have been drawn fron LEDA databases \cite{Paturel03}, absolute B
magnitude has been computed adopting H$_0$ = 70).Moreover, the two
samples have the same fraction (75\%) of triplets.  We found emission
lines in the 67\% of the galaxies of our sample. UZC catalogue
provides a galaxy classification which should indicate the presence of
emission (E), absorption (A) or both (B) feautures in the spectrum.
We found emission lines in all galaxies classified as E or B but also
in 131 of the 344 spectra classified as A.

\section{Nuclear Classification}

The classification of the nuclear activity has been made mainly using
three diagnostic diagrams: 1) log([OIII]5007\AA/H$_{\beta}$) versus
log([NII]6584\AA/ H$_{\alpha}$) or [NII]-Diagram, 2) 
log([OIII]5007\AA)/H$_{\beta}$) versus
log([SII]6717\AA+6731\AA/H$_{\alpha}$) or [SII]-Diagram and 3) 
log([OIII]5007\AA/H$_{\beta}$) vs log([OI]6300\AA/H$_{\alpha}$)
 also called [OI]-Diagram, when [OI] was available. Previous works
(\cite{Kauffmann03}, \cite{Kauffmann04}, \cite{Kewley06}) have shown
the importance of the [NII]-Diagram to do the classification and even,
when [OIII] is not detectable, it has been shown \cite{Stasinska06}
that the use of the ratio [NII](6584\AA)/H$_{\alpha}$ is enough .

\begin{figure}
 \includegraphics[width=9cm]{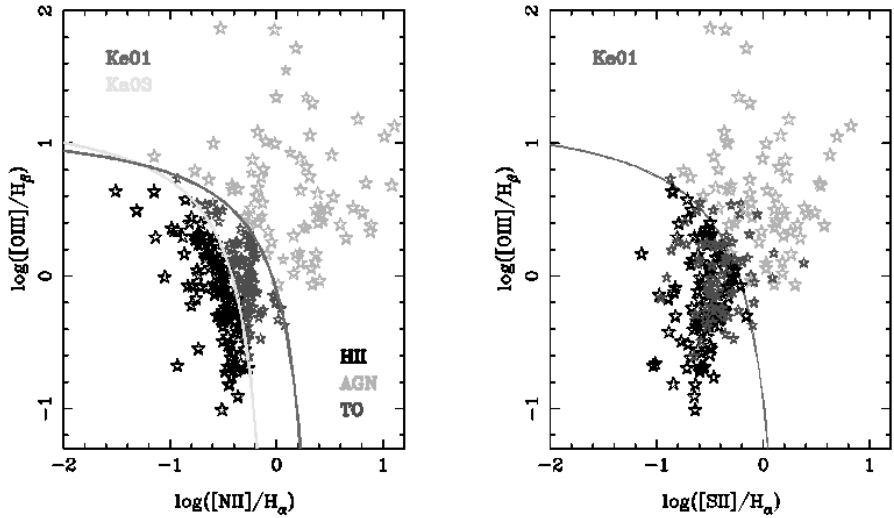}
 
  \caption{Diagnostic diagrams to classify nuclear activity. Clear and
  dark gray lines correspond to Ka03 and Ke01 sequences.}  

    \label{fig3}
    \end{figure}

In Fig \ref{fig3} we show the two "main" diagnostic diagrams for the
galaxies that have measured fluxes for the the six lines. The lines
correspond in light grey to Kauffmann et al. 2003
(hereafter Ka03) sequence for HII nuclei:
\begin{center}
log[OIII]/H$_{\beta}$=(0.61/(log[NII]/H$_{\alpha}$)-0.05)+1.3\\
\end{center}
the dark grey lines to the Kewley et al. 2001(Ke01) sequences:
\begin{center}
log[OIII]/H$_{\beta}$=(0.61/(log[NII]/H$_{\alpha}$)-0.47)+1.19 in NII-diagram\\
log[OIII]/H$_{\beta}$=(0.72/(log[SII]/H$_{\alpha}$)-0.32)+1.30 in SII-diagram\\
\end{center}

We could not detect all emission lines in every spectra,so to classify
nuclear activity we adopted the criteria illustrated below.
\begin{itemize}
\item For 260 galaxies it was possible to use both diagnostic diagrams of 
Fig.\ref{fig3}  thus we classified as:

\begin{itemize}
\item HII nuclei, the galaxies located below Ka03 and below Ke01 respectively 
in the [NII] and [SII] Diagram.
\item AGNs, the galaxies above Ka03 and above Ke01 in the [NII] and 
[SII] Diagram.
\item TOs, the galaxies lying between Ke01 and Ka03 in [NII]-Diagram 
and below Ke01 in [SII]-Diagram.
\end{itemize}

\item For 151 galaxies we could compute only the log([NII]/H$_{\alpha}$) ratio thus
 we classified them as:
\begin{itemize}
\item HII if log([NII]/H$_{\alpha}$) $<$ -0.4 
\item AGN if log([NII]/H$_{\alpha}$) $>$ -0.1
\item classification of galaxies between these two limits was made 
by visual inspection of spectra looking for the presence of
H$_{\beta}$ in emission or/and absorption and the detection of broad
lines.
\end{itemize}

\item Galaxies showing  only [NII] feature were classified
as AGNs according to the study made by \cite{Coziol04}.

\item The remaining 6 galaxies without any of the previous ratios available
were classified on the basis of the detectable emission lines, the
stellar spectral features or the presence of broad lines.
\end{itemize}
As a result of nuclear classification we found:
180 HII, 210 AGNs and 96 T0s, corresponding respectively to 37\%,
43\% and 20\% of galaxies with emission line spectra. The fraction of TOs
is rather large and indicates a high proportion of LLAGNs (Seyfert 2 and
LINERs) in UZC-CGs.

\section{Nuclear activity and host galaxy}
 
\begin{figure}
\centering
\includegraphics[height=12cm,angle=-90]{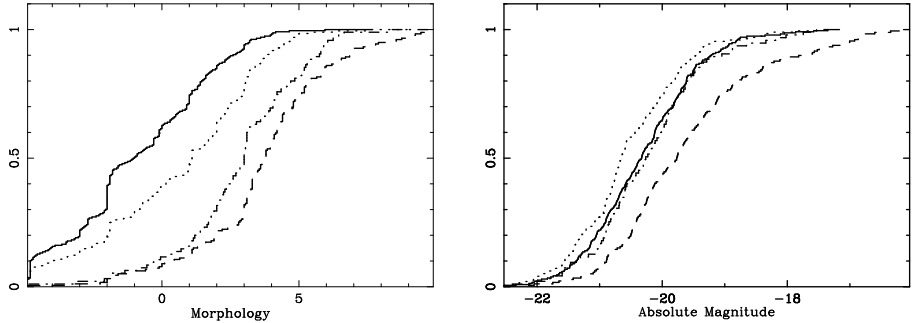}
\caption{Nuclear activity in relation with the properties of the host
galaxy. Solid line correspond to non emission (NoE) galaxies, dotted line to
AGNs, dashed line to HII and dotted-dashed line to Transition Objects (TOs).}
\label{fig4}      
\end{figure}

We have looked for possible relation between nuclear activity kind and
morphological and photometrical properties of the host galaxy.  In
table \ref{table2} we show the median values of morphology, radial
velocity and absolute magnitude (M$_B$) of galaxies hosting different
kind of nuclear activity.  In figure \ref{fig4} we can see that AGNs
(dotted line) are located in bright early types and early spirals, HII
(dashed line) are hosted in fainter and later types, while galaxies
hosting TOs (dotted-dashed line) share properties with AGN and HII
host galaxies, in particular they are located in late spirals having
high luminosity. Non emission line galaxies are found in the earliest
galaxy types.
\begin{table}
\caption{Median values of host galaxy characteritics}
            
\label{table2}     
\centering                        
\begin{tabular}{l c c c c}        
\hline\hline               
Classification~ & ~Morphology~ & ~Velocity(km/s)~ & ~Absolute Magnitude \\   
\hline                        
NonE         & -1.0(SO)  & 5887  & -20.3   \\
AGNs         &  1.1(Sa)  & 5628  & -20.6   \\
TO           &  3.0(Sb)  & 5408  & -20.3   \\
HII          &  3.8(Sbc) & 4367  & -19.8   \\
\hline                                   
\end{tabular}
\end{table}

The analysis of the incidence of nuclear activity as a function of
host galaxy morphology shows that 80\% of activity found in early-type
galaxies (E+SOa) is AGN. This fraction rises to 92\% considering
elliptical galaxies only.  In Early Spirals (Sa-Sbc) AGN is still the
predominant kind of activity but the incidence of HII and TOs is large
too.Almost all Late Spirals ($>$Sbc) have emission in their nucleus
which derives from Star Formation phenomena. The relationship between
nuclear activity kind and morphology is thus confirmed in our sample.

\section{Nuclear activity and host groups}

We have investigated possible links between nuclear activity kind and
CG dynamical properties.In table \ref{table3} we list median values of radial velocity, mean
pairwise separation ($R_{p}$) and velocity dispersion ($\sigma_{v}$)
for CGs dominated by AGNs (i.e. more than the 50\% of CG members host
an AGN), by AGNs+TOs, by HII-nuclei and by non emission (NoE)
galaxies.  AGN dominated CGs tend to show a larger velocity dispersion
and smaller size than CGs dominated by any other kind of activity. HII
dominated CGs are the nearest, are larger and have a significantly
lower velocity dispersion.
 \begin{table}
\caption{Dynamical properties (median values) of CGs dominated by different
kind of nuclear activity}             
\label{table3}     
\centering                        
\begin{tabular}{l c c c c}        
\hline\hline               
Dominated by  & v$_r$(km/s) & R$_p$(kpc)  & $\sigma_{v}$(km/s)  \\   
\hline                       
NoE         & 6468   & 80   & 183  \\
AGNs        & 5569   & 58   & 181  \\
AGNs+TO     & 5892   & 60   & 175  \\
HII         & 3918   & 74   & 100  \\
\hline                       
\end{tabular}
\end{table}

%
%
%
\input{referenc}



\printindex
\end{document}

%% file: referenc.tex
%
%

%
%